# *Ab initio* Phase Diagram of $Ta_2O_5$

Yan Gong [1,2,#], Huimin Tang [3,#], Yong Yang [1,2,*], and Yoshiyuki Kawazoe[4,5]

[1]*Key Lab of Photovoltaic and Energy Conservation Materials, Institute of Solid State Physics, HFIPS, Chinese Academy of Sciences, Hefei 230031, China*

[2]*Science Island Branch of Graduate School, University of Science and Technology of China, Hefei 230026, China*

[3]*College of Physics and Technology, Guangxi Normal University, Guilin 541004, China*

[4]*New Industry Creation Hatchery Center, Tohoku University, Sendai, 980-8579, Japan*

[5]*Center for Interdisciplinary Research, SRM University-AP, Neerukonda, Mangalagiri Mandal, Guntur District, Andhra Pradesh, 522240, India*

Tantalum pentoxide ($Ta_2O_5$) is a polymorphic wide-bandgap semiconductor with outstanding dielectric properties and widespread use in optical and electronic technologies. Its rich structural diversity, arising from multiple polymorphs accessible under different synthesis conditions, has made $Ta_2O_5$ a long-standing subject of interest. However, a unified understanding of the thermodynamic stability and phase transitions of its polymorphs across pressure–temperature (*P*-*T*) space has remained elusive. Here, using first-principles calculations, we map the thermodynamic landscape of $Ta_2O_5$ and establish a comprehensive *P*-*T* phase diagram together with a phase-stability hierarchy. We find that γ-$Ta_2O_5$ and B-$Ta_2O_5$ dominate the phase diagram over a broad range of *P*-*T* conditions: γ-$Ta_2O_5$ is stabilized at low pressures, while B-$Ta_2O_5$ becomes thermodynamically favored at higher pressures up to ~ 60 GPa, beyond which Y-$Ta_2O_5$ emerges as the most stable phase. Crucially, the zero-point energy (ZPE), one aspect of nuclear quantum effects (NQEs), plays a significant role in determining relative phase stability, contributing substantially to the Gibbs free energy and altering phase boundaries. A re-entrant phase transition between γ- and B-$Ta_2O_5$ is predicted near ~ 2 GPa, revealing unexpected complexity in the phase behavior of this oxide. More generally, we identify a characteristic temperature ($T_0$), at which zero-point and thermal phonon contributions to the free energy become comparable, and show that $T_0$ is approximately one-third of the Debye temperature. This relationship provides a simple, physically transparent criterion for assessing the importance of NQEs in phase stability, with implications extending beyond $Ta_2O_5$ to a broad class of complex oxides.

*Corresponding Author: yyanglab@issp.ac.cn

[#] These authors contribute equally to this work.

## I. Introduction

Tantalum pentoxide ($Ta_2O_5$) is a versatile and critical material in modern science and engineering, due to its wide bandgap, high dielectric constant, and remarkable chemical stability. These properties have enabled its widespread application in optoelectronic devices [1-4], resistive random-access memory (RRAM) [5-10], and catalysis [11-13].

In optics, $Ta_2O_5$ is used as an anti-reflective coating in photovoltaic devices [14,15] and as a promising material for ultraviolet photodetectors [16,17]. In electronics, its high dielectric constant supports the miniaturization of capacitors [18,19] and dielectric materials [20,21]. Additionally, $Ta_2O_5$ is employed in catalysis for hydrogen production [22-24] and as a coating material in biomedical engineering to enhance biocompatibility and corrosion resistance [25-27]. These unique properties position $Ta_2O_5$ as a promising material for cutting-edge technologies in renewable energy and advanced medical devices.

A thorough understanding of $Ta_2O_5$ polymorphism is essential to fully harness its functional properties. Research in this area can be divided into two main avenues: experimental observation and theoretical understanding. In 1952, Lagergren and Magnéli reported a high-temperature $Ta_2O_5$ phase, providing one of the earliest experimental glimpses into the Ta-O system and laying the groundwork for subsequent structural investigations [28]. Throughout the 1960s, systematic powder X-ray diffraction (XRD) studies revealed multiple orthorhombic and hexagonal polymorphs, establishing a foundation for the rational exploration of $Ta_2O_5$ crystal chemistry [29,30]. Based on this experimental data, Stephenson and Roth proposed the $L_{SR}$ model in 1971-an orthorhombic structural model containing 11 formula units (22 Ta/55 O)-though it required oxygen vacancies to maintain stoichiometry [31]. Researchers subsequently developed simplified versions of the $L_{SR}$ model, including the $\beta_{AL}$ model by Aleshina and Loginova [32] and the $\beta_R$ model by Ramprasad [33]. High-pressure experiments discovered the B and Z phases [34], while thin-film deposition techniques such as chemical vapor deposition and magnetron sputtering successfully prepared the hexagonal δ-phase [35]. In 1992, Hummel *et al.* reported the synthesis of TT-$Ta_2O_5$ and determined via X-ray diffraction its lattice parameters ($a$ = 3.639 Å, $c$ = 3.901 Å) and confirmed its P6/*mmm* space group, though the precise atomic structure remained unresolved [36]. By the 21st century, high-resolution synchrotron radiation and electron microscopy experiments had progressively refined the $Ta_2O_5$ polymorph database, providing reliable references for theoretical studies.

On the theoretical side, first-principles calculations and structure search algorithms played pivotal roles in elucidating the stability of $Ta_2O_5$ polymorphs. In 1997, Fukumoto and Miwa used density functional theory (DFT) to precisely resolve the atomic structure of the δ-phase [37] for the first time, confirming its P6/*mmm* space group and strong agreement with experimental diffraction patterns [36]. Later, Lee *et al*. proposed the orthorhombic λ-phase (Pbam space group) and demonstrated its lower energy (higher stability) than the $L_{SR}$ model [38]. In 2016, Perez-Walton *et al.* systematically compared the thermodynamic and kinetic stabilities of multiple $Ta_2O_5$ phases, finding that those composed of $TaO_6$ octahedra (e.g., B, λ, $\beta_{AL}$) exhibited the lowest formation energies and dynamical stability, whereas structures incorporating higher-coordination polyhedra (e.g., $TaO_7$, $TaO_8$)-such as $L_{SR}$, $L_G$, T, H, δ, and Z-were either higher in energy or dynamically unstable, existing only under metastable or defective polymorphs [39]. In 2018, Yang and Kawazoe employed *ab initio* method and evolutionary algorithms to predict a new stable phase (γ-phase) [40] consisting entirely of ordered, distorted $TaO_6$ octahedral networks, establishing it as the most plausible ground-state theoretical model at low temperatures [41,42]. Further, in 2023, Yang's group combined particle swarm optimization (PSO) and DFT calculations to identify the stable triclinic $\gamma_1$-$Ta_2O_5$ under ambient pressure, whose unit cell contains only one formula unit (Z = 1) [43].

While previous theoretical studies have focused on predicting the possible new individual $Ta_2O_5$ phases, a comprehensive understanding of their interrelationships (e.g., mutual transitions between every two phases) requires thermodynamic and/or kinetic analyses. Generally, the relationships among different crystal phases of a compound can be established through phase diagrams. The study of phase diagrams is crucial for revealing the dependence of material structures on temperature and pressure, providing precise temperature-pressure ($T$-$P$) windows for experimental synthesis, and predicting material behavior under extreme conditions, thereby bridging theoretical predictions with practical applications. Brennan *et al.* systematically compared various proposed $Ta_2O_5$ structures using room-temperature high-pressure synchrotron XRD and DFT calculations, identifying seven potential phase models for the powder samples of $Ta_2O_5$ [44]. Their analysis revealed consistent equation-of-state parameters ($K_0 = 138 \pm 3.68$ GPa, $K_0^{'} = 1.82 \pm 0.45.$) across all candidate phases, and identified the λ-phase to be the best structural model for the $Ta_2O_5$ samples employed in their study. They also confirmed the pressure-induced amorphization of $Ta_2O_5$ at approximately 25 GPa, which shows remarkable consistency with the critical amorphization pressure reported in prior studies [45,46], thereby reinforcing the reproducibility of this phase transition behavior. To further resolve structural uncertainties, they conducted neutron diffraction experiments on $Ta_2O_5$ powders at temperatures up to 1000 ºC [47]. The enhanced sensitivity of neutron diffraction to oxygen positions in the presence of heavy Ta atoms provided evidence that an orthorhombic $L_{SR}$-type structure best explains the experimental data, outperforming other theoretical models. In spite of these progresses, the lack of simultaneous high-pressure studies means a complete pressure-temperature ($P$–$T$) phase diagram remains unestablished [44,47]. Recently, we predicted a potential orthorhombic phase, termed Y-$Ta_2O_5$, under high-pressure conditions using DFT calculations combined with structure-searching algorithms [48]. This phase contains four formula units per unit cell (Z = 4) and exhibits the highest known Ta–O coordination number among all reported $Ta_2O_5$ polymorphs. This study revealed that Y-$Ta_2O_5$ is the most energetically favorable phase of $Ta_2O_5$ within the pressure range of approximately above 60 GPa to at least 200 GPa [48]. Despite these advances, a complete phase diagram across wide temperature-pressure ranges remains lacking.

Building upon previous studies, this paper systematically evaluates the thermodynamic stability of various $Ta_2O_5$ polymorphs under a wide range of temperature and pressure conditions using first-principles calculations. A comprehensive equilibrium phase diagram and a phase stability order diagram are established, offering a robust theoretical data for understanding phase transitions in $Ta_2O_5$ polymorphs.

Another focus is on the nuclear quantum effects (NQEs) [49-58]. It is shown that one aspect of NQEs, namely the zero-point energy (ZPE), together with the thermal atomic vibrations (quantization as phonons)[59], can induce significant modifications to the phase diagram and phase stability sequence. This observation demonstrates that NQEs remain significant even in systems lacking of light elements (e.g., hydrogen, helium). Further analysis revealed that the ZPE equals to the excited vibration energy at a characteristic temperature $T_0$, which is found to be approximately one-third of the Debye temperature.

The contents of this paper are organized as follows: After the Introduction part, Section II introduces the theoretical method employed in this study and the technical details of the DFT calculations. Section III presents the *ab initio* results of phase diagram and phase stability diagram of $Ta_2O_5$ polymorphs under a wide range of temperature—pressure conditions, and analyzes in general the physical consequences due to nuclear quantum effects (NQEs), in particular the role of zero-point energy (ZPE). The main conclusion is presented in Section IV.

## II. Methods

The Vienna Ab Initio Simulation Package (VASP) based on DFT was employed for structural optimization and total energy calculations [60,61]. The Kohn-Sham equations were solved using a plane-wave basis set with an energy cutoff of 600 eV, and the projector-augmented wave (PAW) method was used to describe ion-electron interactions [62,63]. The Perdew-Burke-Ernzerhof (PBE) [64] type generalized gradient approximation (GGA) was used to describe the exchange-correlation interactions of electrons. The structures of the $\gamma$, $\gamma_1$, B, $\lambda$, $L_{SR}$, $\delta$, $\beta_{AL}$, $\beta_R$, Z and Y phases of $Ta_2O_5$ were optimized, and their total energies were calculated as a function of unit cell volume. The Gibbs free energy were calculated with the aid of the PHONOPY package [65,66], which accounts for both static electronic and phonon vibrational contributions based on density functional perturbation theory (DFPT) [67-69]. By offering a reliable foundation for comparing phase stability under different temperature and pressure conditions, it facilitates the precise construction of P-T phase diagram. In case of structural relaxation and total energy calculations, the following k-point meshes generated using the Monkhorst-Pack scheme [70] were used for Brillouin zone (BZ) sampling of the indicated phase: 8 × 8 × 2 for γ, 8 × 8 × 6 for $\gamma_1$, 2 × 4 × 4 for B, 4 × 4 × 8 for λ, 4 × 2 × 4 for $L_{SR}$, 4 × 4 × 8 for δ, 4 × 8 × 4 for $\beta_{AL}$, 4 × 4× 8 for $\beta_R$, 4 × 4 × 4 for Z, and 2 × 4 × 4 for Y. In the DFPT calculations, the k-point meshes were uniformly reduced by a factor of two along each reciprocal lattice direction. Accordingly, 4 × 4 × 1 for γ, 4 × 4 × 3 for $\gamma_1$, 1 × 2 × 2 for B, 2 × 2 × 4 for λ, δ, and $\beta_R$, 2 × 1 × 2 for $L_{SR}$, 2 × 4 × 2 for $\beta_{AL}$, 2 × 2 × 2 for Z and 1 × 2 × 2 for Y were used. To verify the stability of γ and B phases, *ab initio* molecular dynamics (AIMD) simulations were further performed. The simulations were carried out in canonical ensembles (NVT) with the temperature controlled by a Nosé–Hoover thermostat [71-73]. For each temperature, an initial 1 picosecond (ps) run was used for thermal equilibration, followed by a 5 ps production run to ensure statistical convergence. A time step of 1 femtosecond (fs) was used throughout. The finite-temperature dynamical stability of $Ta_2O_5$ was evaluated by analyzing the evolution of total energy and atomic configurations over time.

## III. Results and Discussions

### A. Constitutive Properties

The crystal unit cells of the ten phases of $Ta_2O_5$ (γ, $\gamma_1$, B, $L_{SR}$, $\beta_{AL}$, λ, δ, $\beta_R$, Z, and Y) are schematically shown in Fig. 1. The crystal structure of $Ta_2O_5$ is primarily composed of $TaO_n$ polyhedra, with Ta atoms at the centers and O atoms at the vertices [33,74,75]. Based on the number of oxygen atoms bonded to the central Ta atom, the polyhedra can be classified into four types: octahedron ($TaO_6$), pentagonal bipyramid ($TaO_7$), hexagonal bipyramid ($TaO_8$) and $TaO_{10}$ polyhedra. Specifically, the $\beta_{AL}$, $\beta_R$, B, λ, γ, and $\gamma_1$ phases are composed of distorted octahedra; the hexagonal δ phase consists of edge-sharing octahedra and hexagonal bipyramids; the $L_{SR}$ phase consists of corner-sharing octahedra and pentagonal bipyramids; the Z phase is composed solely of pentagonal bipyramids; and the Y phase consists of $TaO_{10}$ polyhedra (Fig. 1). The optimized lattice parameters for the ten phases of $Ta_2O_5$ are presented in Table I. Notably, the Y phase only exists stably above 62 GPa, which can be viewed as the transformation of the B phase under such high-pressure condition.

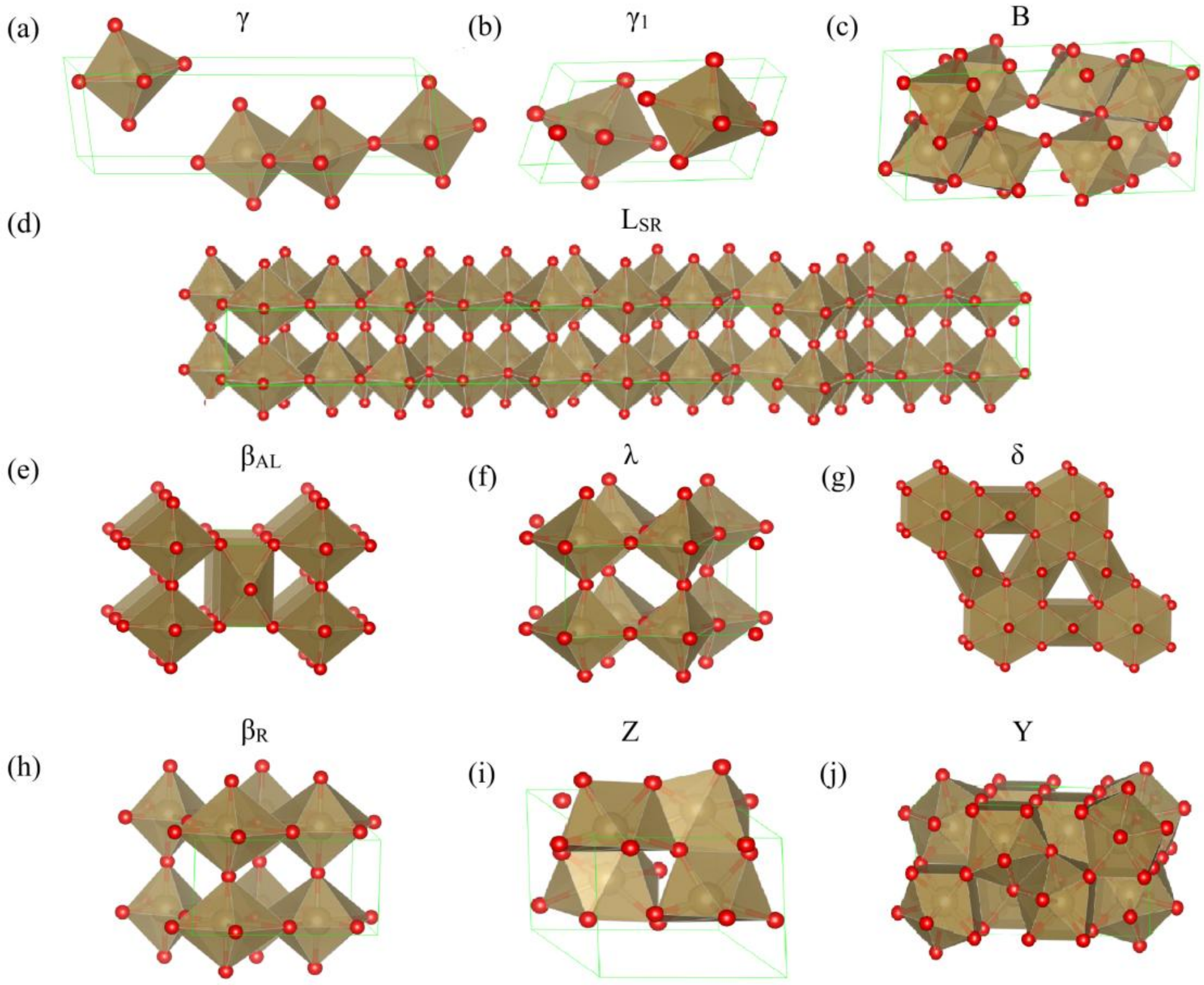

**FIG. 1.** Schematic diagram of the structure models of $Ta_2O_5$: (a) γ; (b) $\gamma_1$; (c) B; (d) $L_{SR}$; (e) $\beta_{AL}$; (f) λ; (g) δ; (h) $\beta_R$; (i) Z and (j) Y. The O and Ta atoms are represented by smaller (red) and larger (brown) spheres, respectively.

**Table I.** The calculated lattice parameters of the ten $Ta_2O_5$ phases, where *α*, *β*, and *γ* represent the angles between the cell edges *b* and *c*, *a* and *c*, and *a* and *b*, respectively. Z represents the number of $Ta_2O_5$ formula units in the unit cell.

| $Ta_2O_5$ | *a* (Å) | *b* (Å) | *c* (Å) | *α* (°) | *β* (°) | *γ* (°) | Z |
|---|---|---|---|---|---|---|---|
| γ | 3.89 | 3.89 | 13.38 | 81.77 | 98.25 | 89.98 | 2 |
| $\gamma_1$ | 3.87 | 3.91 | 6.82 | 90.35 | 73.49 | 90.00 | 1 |

| B | 12.93 | 4.92 | 5.59 | 90.00 | 103.23 | 90.00 | 4 |
|---|---|---|---|---|---|---|---|
| $L_{SR}$ | 6.33 | 40.92 | 3.85 | 90.00 | 90.00 | 89.16 | 11 |
| $\beta_{AL}$ | 6.52 | 3.69 | 7.77 | 90.00 | 90.00 | 90.00 | 2 |
| λ | 6.25 | 7.40 | 3.82 | 90.00 | 90.00 | 90.00 | 2 |
| δ | 7.33 | 7.33 | 3.89 | 90.00 | 90.00 | 120.00 | 2 |
| $\beta_{R}$ | 7.42 | 6.21 | 3.87 | 90.00 | 90.00 | 90.00 | 2 |
| Z | 6.05 | 5.19 | 5.82 | 90.00 | 118.33 | 90.00 | 2 |
| Y | 9.94 | 4.98 | 4.32 | 90.00 | 90.00 | 90.00 | 4 |

**B. Phase Diagrams, Phase Stability Order Diagrams, and NQEs**

In many-particle systems such as condensed phase materials, the Gibbs free energy is the fundamental thermodynamic quantity that dictates phase stability, phase transitions, and the construction of phase diagrams. For semiconductors and insulators, where electrons largely remain in their ground states at finite temperatures, the free energy is dominated by lattice contributions—namely, the internal lattice energy, the phonon free energy, and the pressure-volume term (*PV*). Under temperature *T*, pressure *P* and volume *V*, the Gibbs free energy (*G*) can be expressed as: $G = F + PV$, where $F = -k_B T \ln Z$, and $Z = \exp(-U/k_B T)\prod_j \frac{exp(-\hbar\omega_j/2k_B T)}{[1-exp(-\hbar\omega_j/k_B T)]}$. It follows that:

$$F = U + \frac{1}{2}\sum_j \hbar\omega_j + k_B T \sum_j \ln\left[1-\exp(-\hbar\omega_j/k_B T)\right] \tag{1}$$

where $U$ represents the static free energy or zero-temperature ground state energy, which refers to the system's internal energy when phonons are not taken into account. Generally, it can be obtained directly through DFT calculations. $k_B$ is the Boltzmann constant, and $\hbar$ is the reduced Planck constant. The second term ($\frac{1}{2}\sum_j \hbar\omega_j$) is the zero-point energy (ZPE), which is due to the quantum motions of atoms (more precisely, the atomic nuclei) at absolute zero temperature ($T = 0$), and $\omega_j$ denotes the vibrational frequency of the *j*-th phonon mode. The phonon free energy $F_{ph}(V,T) = \frac{1}{2}\sum_j \hbar\omega_j + k_B T\sum_j \ln\left[1-exp(-\hbar\omega_j/k_B T)\right] = E_{ph} - TS$. The phonon energy $E_{ph} =$

$\sum_j \left(\frac{1}{2}\hbar\omega_j + \frac{\hbar\omega_j}{e^{\hbar\omega_j/(k_B T)}-1}\right)$, and the entropy term due to phonon is given by $TS = \sum_j \frac{\hbar\omega_j}{e^{\hbar\omega_j/(k_B T)}-1} - k_B T \sum_j \ln\left[1 - exp(-\hbar\omega_j/k_B T)\right]$.

These quantities can be readily obtained using VASP jointly with PHONOPY, which yields the Gibbs free energy $G(P,T) = U + F_{ph} + PV$. With the computed Gibbs free energies, the relative appearance probability of each polymorph can be evaluated under given $P$-$T$ conditions. The equilibrium phase diagram of $Ta_2O_5$ is constructed by comparing $G(P,T)$ across nine polymorphs ($\gamma$, $\gamma_1$, B, $\lambda$, $L_{SR}$, $\delta$, $\beta_{AL}$, $\beta_R$, and Z) over a range of pressures and temperatures, identifying the phase with the lowest $G$ at each ($P$, $T$) point, as well as establishing the phase stability order diagram.

Figure 2 shows the temperature dependence of the Gibbs free energy, $G$(T), for nine $Ta_2O_5$ polymorphs under pressures from 0 to 60 GPa, with the panels (a)-(h) representing 0, 6, 10, 20, 30, 40, 50, and 60 GPa, respectively. In all cases, the free energies decrease monotonically with temperature, reflecting the increasing contribution of vibrational entropy. At ambient pressure (panel a), the $\gamma$ and $\gamma_1$ phases are the first two most stable, while the $\beta_{AL}$ and Z phases are the least stable. As pressure rises, the relative free energies change significantly: the B phase becomes thermodynamically favored above 6 GPa, whereas the $\gamma$ and $\gamma_1$ phases gradually lose their thermodynamical stability. Futhermore, the $\lambda$ phase becomes the second most stable high-pressure phase when $P \geq 4.2$ GPa (inset of Fig. 2(b)), and remains so untill ~ 31.8 GPa.

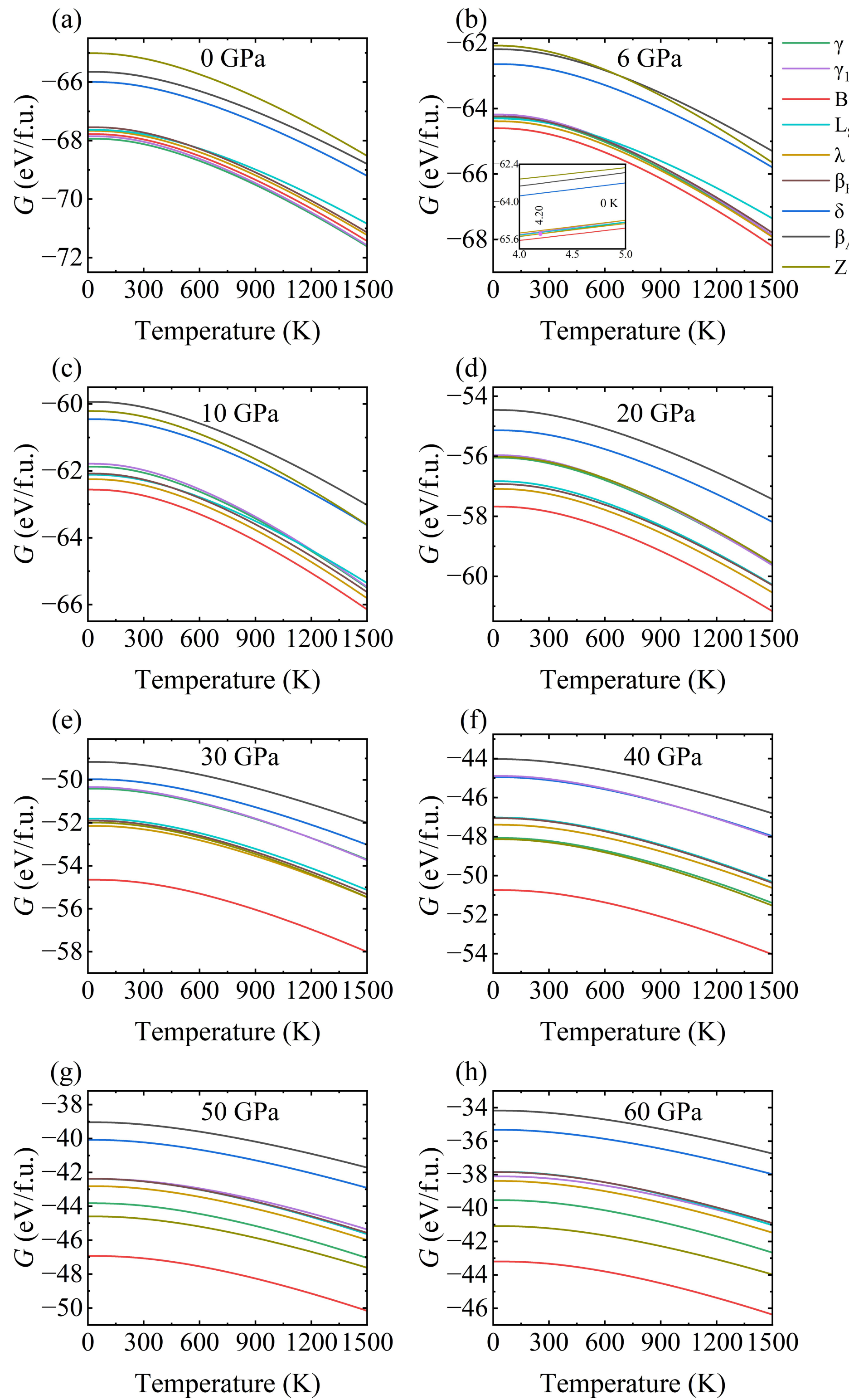

**FIG. 2.** Temperature-dependent Gibbs free energies $G(T)$ of nine $Ta_2O_5$ polymorphs under pressures from 0 to 60 GPa, in units of eV per formula unit (eV/f.u.). Panels (a)-(h) correspond to pressures of 0, 6, 10, 20, 30, 40, 50, and 60 GPa, respectively.

Figure 3 provides an overview of the stability rankings of the nine $Ta_2O_5$ polymorphs over a broad pressure-temperature ($P$-$T$) range, covering temperatures of 0, 600, and 1200 K and pressures from 0 to 60 GPa. At $P$ ~ 0 GPa (ambient pressure), the $\gamma$ and $\gamma_1$ phases are the most and second-most stable, followed by the B and $\lambda$ phases, whereas the Z and $\beta_{AL}$ phases exhibit the lowest stability. With increasing pressure, the stability hierarchy undergoes a pronounced restructuring: the B phase emerges as the most stable polymorph at intermediate and high pressures, while the relative stability of the $\gamma$- and $\gamma_1$-type structures progressively decreases.

As seen from Fig. 3, the influence of temperature on the stability ranking is pressure dependent. At $P$ = 0 GPa, when the temperature increases to $T \geq 600$ K, only the stability order of $L_{SR}$ and $\beta_R$ is reversed, while the ordering of all other phases remains unchanged, and this sequence persists up to 1200 K. At $P$ = 20 GPa, the temperature effect is similarly weak, with a stability reversal occurring only between Z and $\gamma_1$ at $T \geq 600$ K, and no further changes observed at 1200 K. At $P$ = 40 GPa, the stability sequence remains identical from 0 to 600 K, whereas at 1200 K a minor reordering is observed exclusively between $\gamma_1$ and $\delta$. At $P$ = 60 GPa, the temperature effect is similarly weak, with a stability reversal occurring only between $L_{SR}$ and $\beta_R$ at $T \geq 600$ K, and no further changes observed at 1200 K. Together, Fig. 3 illustrates how pressure and temperature change the stability hierarchy of $Ta_2O_5$ polymorphs.

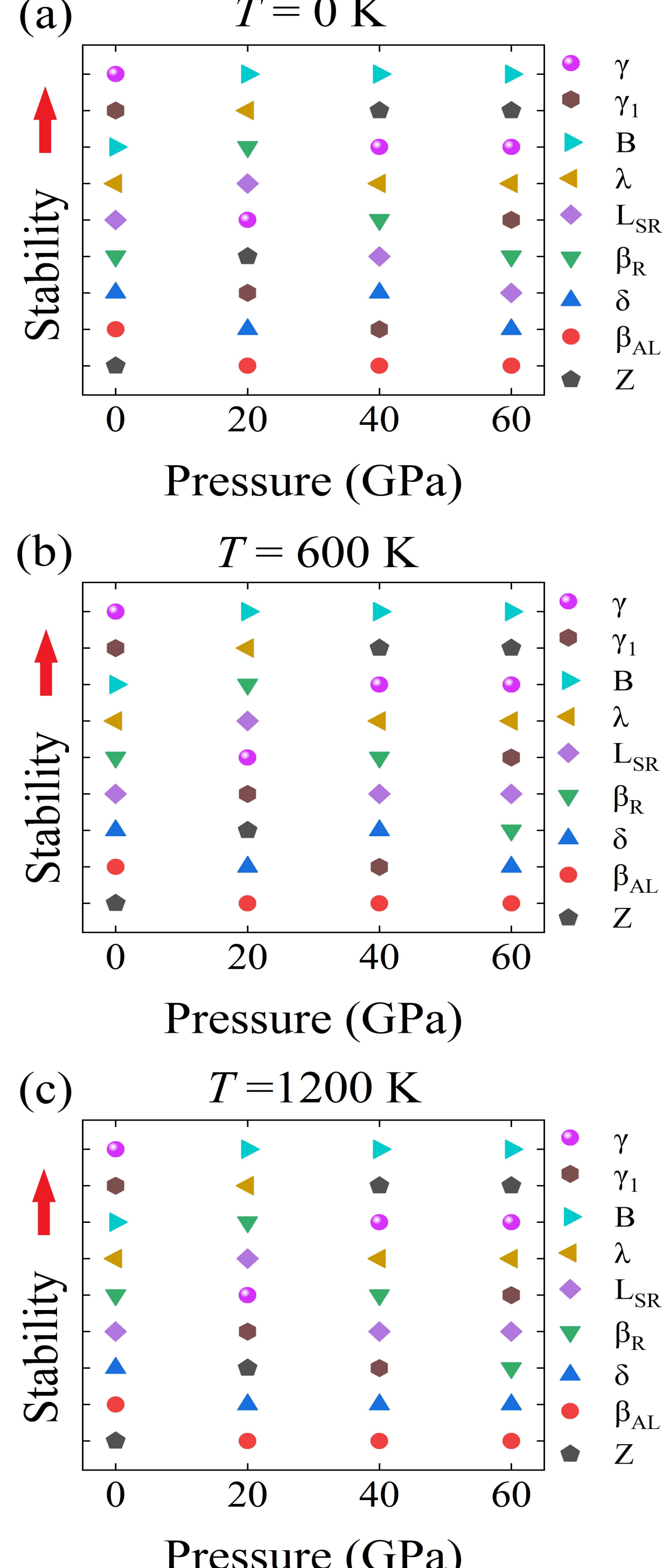

**FIG. 3.** Thermodynamic stability ranking of $Ta_2O_5$ polymorphs, determined from Gibbs free energies, at pressures $P$ = 0, 20, 40, and 60 GPa and temperatures $T$ = 0 K (a), 600 K (b), and 1200 K (c). The arrow indicates increasing stability from bottom to top.

To describe the relative appearance probability of each $Ta_2O_5$ phase, we introduce the quantity $Z_0 = \Sigma_{i=1}^{9} e^{-\Delta G_i/(k_B T)}$, which is similar to the partition function defined in statistical mechanics. Here, $\Delta G_i$ represents the free energy difference of the $i$th phase relative to the minimum of $G(P, T)$. The appearance probability of the $i$th phase is therefore evaluated as follows:

$$\Gamma_i(T, P) = \frac{e^{-\Delta G_i/(k_B T)}}{Z_0} \tag{2}$$

As shown in Fig. 4, the curves illustrate the thermodynamic probabilities of various $Ta_2O_5$ phases under a series of pressures as a function of temperature. Under zero or ambient pressure conditions, the γ phase dominates over a wide temperature range of 0-1300 K, followed by the $\gamma_1$ phase. As temperature increases, the thermodynamic probabilities clearly separate the phases into two distinct groups: The first group consists of six phases (γ, $\gamma_1$, B, λ, $L_{SR}$, and $\beta_R$), while the second comprises three phases (δ, $\beta_{AL}$, and Z). Within each group, the occurrence probabilities of the phases tend to converge to the same value in the high-temperature region.

With increasing external pressure, the B phase rapidly becomes the most thermodynamically stable polymorph, while the stability hierarchy of metastable phases undergoes significant reorganization, splitting them into distinct groups (Figs. 4(b)-(d)). All metastable phases exhibit the same trend: their occurrence probabilities increase substantially with rising temperature. However, under a given external pressure, the relative stability ranking of phases remains essentially unchanged. This indicates that pressure exerts a more pronounced influence on Gibbs free energy than temperature within the range of pressure-temperature investigated. Fundamentally, 1 GPa pressure change induces an energy shift of ~ 6.25 meV per 1 $Å^3$ volume change, which is equivalent to a temperature variation of approximately 72.5 K, which explains the sluggish temperature-dependence of phase stability with comparison to pressure. This also helps to understand the variation trend of phase stability order presented in Fig. 3.

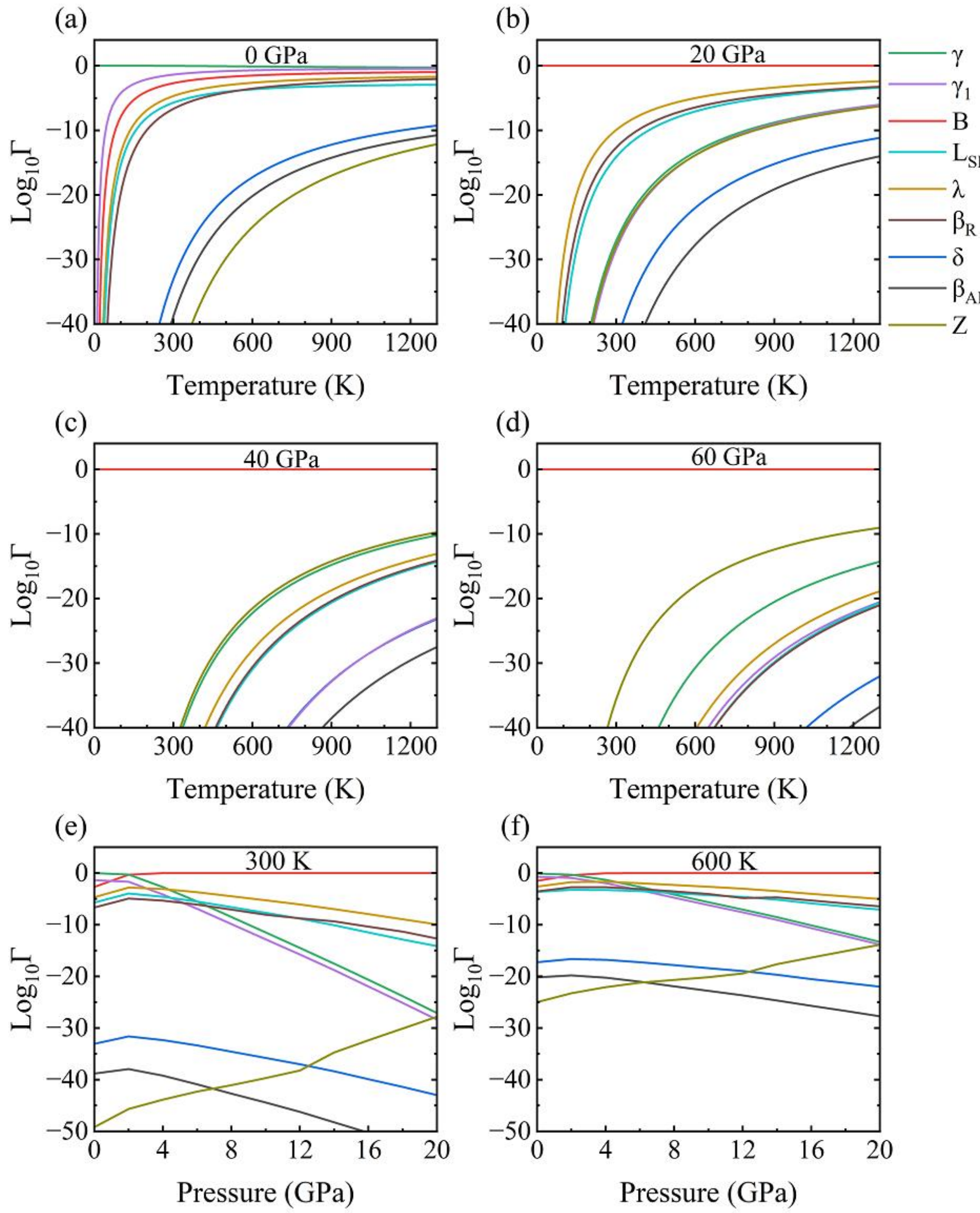

**FIG. 4.** Panels (a) to (d): Variation of the probability of occurrence of nine phases under different pressure conditions, with pressure ranging from 0 GPa to 60 GPa in steps of 20 GPa. Panels (e)-(f): Probability of occurrence of phase at $T$ = 300, 600 K, as a function of pressure.

Based on the analysis above, the Gibbs free energies of nine $Ta_2O_5$ polymorphs over a wide pressure-temperature ($P$-$T$) range were calculated to construct the equilibrium phase diagram. At each ($P$, $T$) point, the stable phase was identified as that with the lowest Gibbs free energy (Fig. 5). By including phonon contributions, we can assess the effects of nuclear quantum effects (NQEs), as considered in this work through zero-point energy (ZPE) which dominates at room temperature and below, and thermal phonon contributions (quantization of collective vibrations of atomic nuclei) [59], on phase stability and transition boundaries, in comparison with the results obtained without such contributions. As seen from Fig. 5(a), when phonon contributions are included, the γ→B phase transition occurs at ~2.0 GPa and 60 K, and the reverse transition (B→γ) at ~2.0 GPa and 208 K. In contrast, under the static lattice approximation [Fig. 5(b)], the B→γ transition shifts dramatically to higher temperatures, occurring at $P$ = 2.0 GPa and $T$ = 571 K. Similarly, the Y→B transition boundary shifts from the ($P$, $T$) point of ~ (62 GPa, 50 K) to ~ (62 GPa, 370 K). These results reveal that NQEs can dramatically change the phase transition temperatures, reshaping the $Ta_2O_5$ phase diagram.

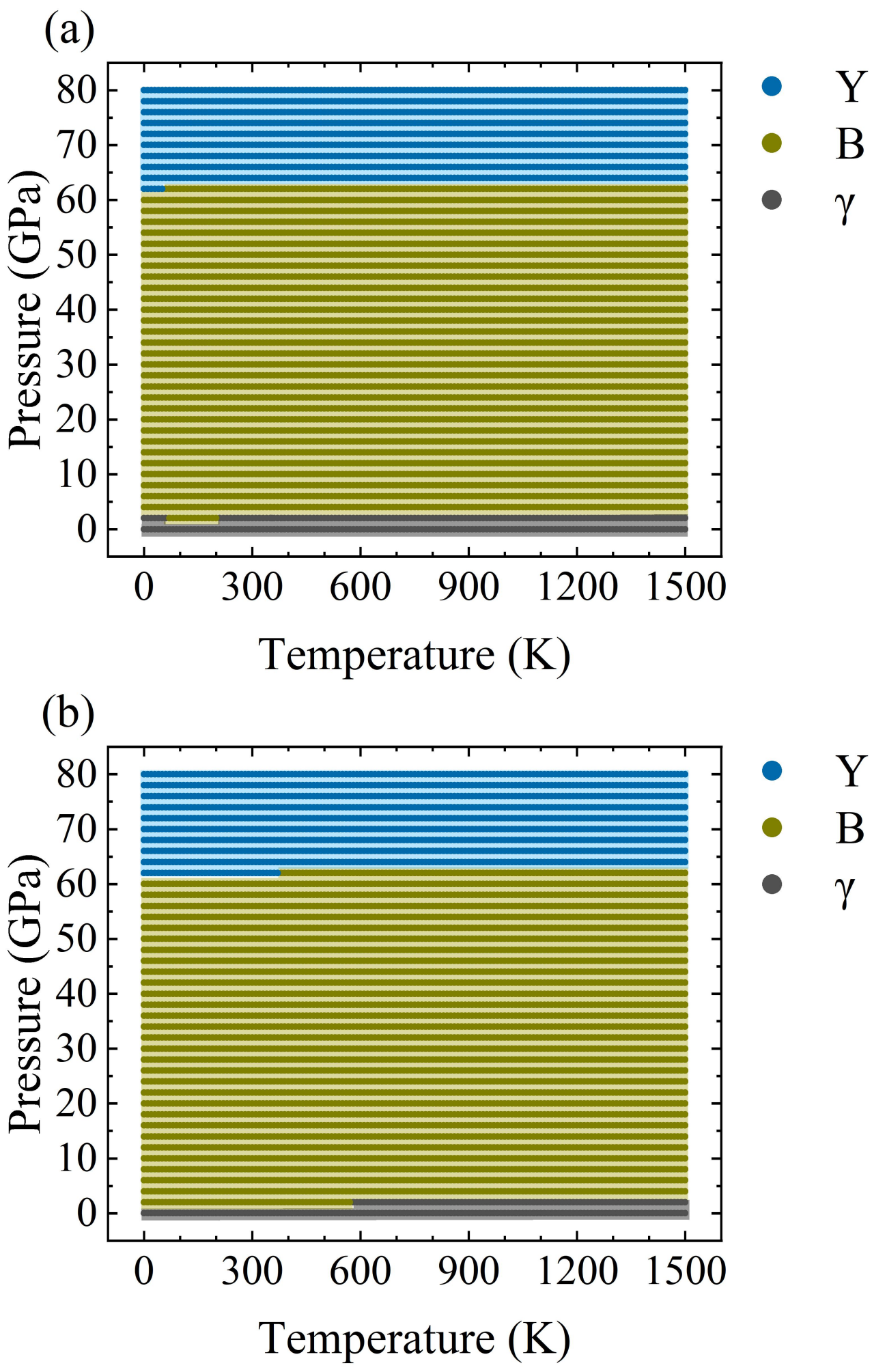

**FIG. 5.** (a) Phase diagram of $Ta_2O_5$ under different temperature and pressure conditions. (b) Temperature–pressure phase diagram of $Ta_2O_5$ within the static lattice

approximation (neglecting phonon contributions).

Table II further illustrates how the inclusion of phonon free-energy contributions ($E_{ph}$) affects the stability ordering of the nine $Ta_2O_5$ polymorphs at $P \sim 0$ GPa (ambient pressure). The results show that incorporating $E_{ph}$ leads to noticeable modifications in the relative stability, particularly at elevated temperatures (e.g., 1500 K), where phonon contributions induce a reordering of the terminal phases of the stability order sequence ($\beta_{AL}$ and Z). This demonstrates that the phonon-derived free-energy term plays a substantive role in reshaping the phase sequence, especially when thermal vibrations become significant. Overall, while the major phases—$\gamma$, $\gamma_1$, B, and $\lambda$—maintain a consistent order across the examined temperature range, the inclusion of $E_{ph}$ clearly alters the relative stability among several metastable phases, highlighting that phonon contributions constitute an essential factor governing the ambient-pressure phase-stability landscape of $Ta_2O_5$.

**Table II.** Temperature-dependence of the order of stability of nine $Ta_2O_5$ polymorphs at pressure $P = 0$.

| Temperature (K) | Order of Stability |
|---|---|
| 0 (without $E_{ph}$) | $\gamma > \gamma_1 > B > \lambda > L_{SR} > \beta_R > \delta > \beta_{AL} > Z$ |
| 0 (with $E_{ph}$) | $\gamma > \gamma_1 > B > \lambda > L_{SR} > \beta_R > \delta > \beta_{AL} > Z$ |
| 300 (without $E_{ph}$) | $\gamma > \gamma_1 > B > \lambda > L_{SR} > \beta_R > \delta > \beta_{AL} > Z$ |
| 300 (with $E_{ph}$) | $\gamma > \gamma_1 > B > \lambda > L_{SR} > \beta_R > \delta > \beta_{AL} > Z$ |
| 600 (without $E_{ph}$) | $\gamma > \gamma_1 > B > \lambda > \beta_R > L_{SR} > \delta > \beta_{AL} > Z$ |
| 600 (with $E_{ph}$) | $\gamma > \gamma_1 > B > \lambda > \beta_R > L_{SR} > \delta > \beta_{AL} > Z$ |
| 900 (without $E_{ph}$) | $\gamma > \gamma_1 > B > \lambda > \beta_R > L_{SR} > \delta > \beta_{AL} > Z$ |
| 900 (with $E_{ph}$) | $\gamma > \gamma_1 > B > \lambda > \beta_R > L_{SR} > \delta > \beta_{AL} > Z$ |
| 1200 (without $E_{ph}$) | $\gamma > \gamma_1 > B > \lambda > \beta_R > L_{SR} > \delta > \beta_{AL} > Z$ |
| 1200 (with $E_{ph}$) | $\gamma > \gamma_1 > B > \lambda > \beta_R > L_{SR} > \delta > \beta_{AL} > Z$ |
| 1500 (without $E_{ph}$) | $\gamma > \gamma_1 > B > \lambda > \beta_R > L_{SR} > \delta >$ **$Z > \beta_{AL}$** |
| 1500 (with $E_{ph}$) | $\gamma > \gamma_1 > B > \lambda > \beta_R > L_{SR} > \delta >$ **$\beta_{AL} > Z$** |

Then we extended the datasets illustrated in Fig. 3, systematically analyzing the thermodynamic stability sequence of different $Ta_2O_5$ polymorphs under a set of temperature-pressure conditions. Using a pressure and temperature increment of $\Delta P$ = 2 GPa and $\Delta T$ = 100 K, Fig. 6 quantifies phase stability evolution across a broad temperature-pressure conditions, directly comparing the cases with/without phonon contributions. The phase stability order at a given $P$-$T$ condition is then determined. For instance, at ambient pressure (0 GPa, 0 K), the stability sequence follows $\gamma > \gamma_1 > B > \lambda > L_{SR} > \beta_R > \delta > \beta_{AL} > Z$. A comparison between Fig. 6(a) and Fig. 6(b) reveals that the stability sequence is significantly altered, underscoring the critical role of phonon contributions and NQEs in refining the phase boundaries and stability hierarchy.

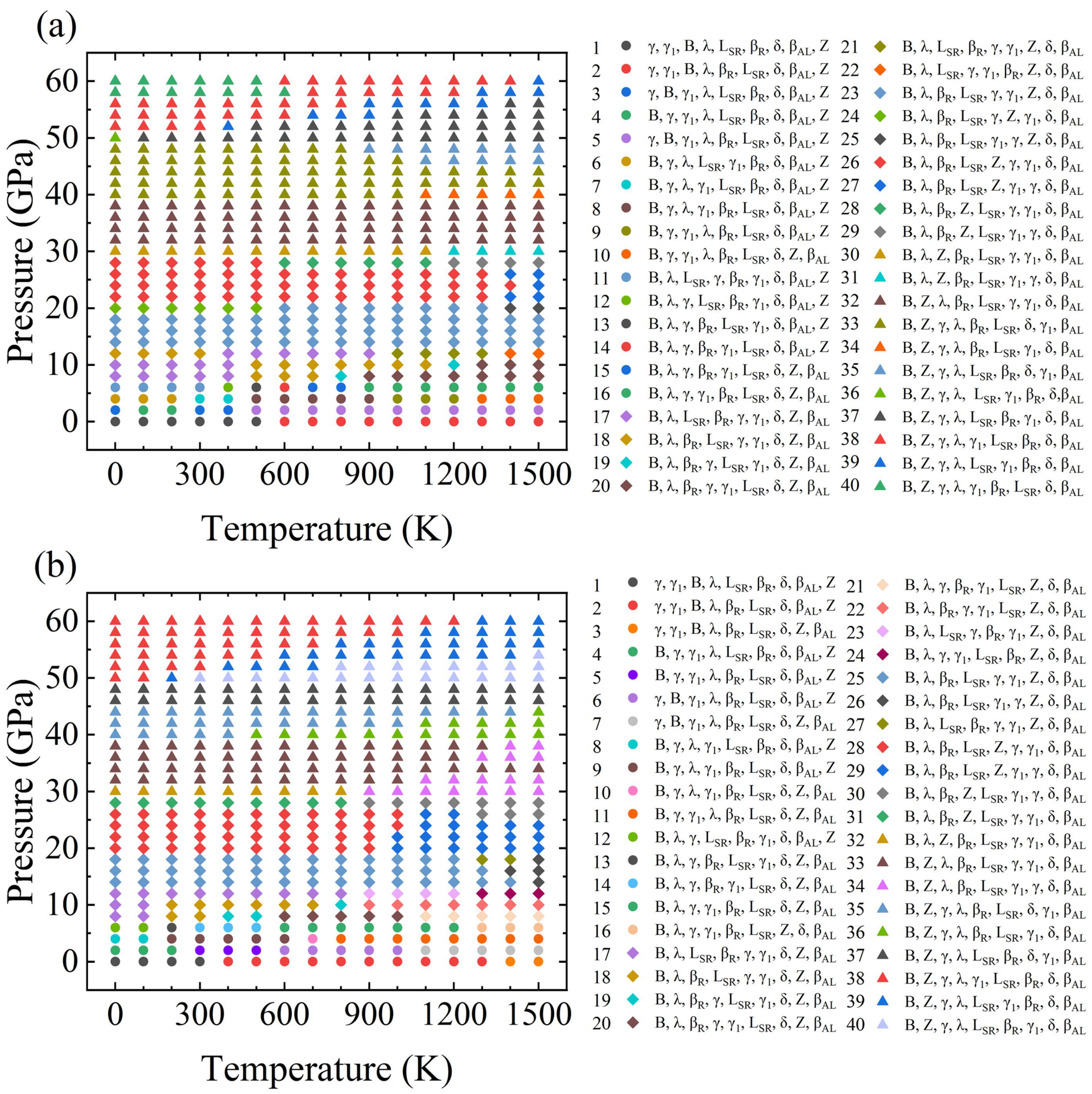

**FIG. 6.** (a) Stability sequence of each phase at a pressure variation step of $\Delta P$ = 2 GPa and a temperature step of $\Delta T$ = 100 K. (b) Stability sequence of various $Ta_2O_5$ polymorphs calculated with a pressure step of 2 GPa and a temperature step of 100 K,

under the static lattice approximation (excluding phonon contributions).

To further refine the analysis, we conducted detailed calculations in the vicinity of the first intersection point in the phase diagrams for both scenarios with and without phonon contributions. This allows us to accurately determine the phase transition pressures ($P_c$) and temperatures ($T_c$). As shown in Fig. 7(a), when phonon energy contributions are taken into account, the γ → B transition of $Ta_2O_5$ occurs at $P_c$ = 2.0 GPa and $T_c$ = 60 K. Upon further heating, the B phase at the same pressure reversely transforms back to the γ phase at approximately 208 K. Such a non-monotonic stability sequence—"stable phase → metastable phase → stable phase" under fixed pressure—is a hallmark of re-entrant phase transition behavior. Re-entrant transitions have been observed or predicted in some materials, including the triangular antiferromagnetic re-entrance in the antiperovskite $Mn_3ZnN$ [76], the α → γ → α non-monotonic sequence in solid nitrogen at ultralow temperatures and high pressures [77], the II → III → II re-entrant structural progression of $CF_4$ under compression [78], the metal–semiconductor–metal electronic re-entrance in lithium [79], as well as symmetry-cycling transitions such as SmC–SmA–SmC in various molecular crystals and liquid-crystal systems driven by enthalpy–entropy competition [80,81]. To our knowledge, this work is the first report on the re-entrant transition in transition metal oxide. In contrast, Fig. 7(b) presents the results without phonon contributions, where the γ-to-B phase transition takes place under single *P*-*T* points: 1.8 GPa at 167 K, 1.9 GPa at 340 K, and 2.0 GPa at 571 K. The re-entrant transition is clearly absent, which demonstrates again the significant role of NQEs in modifying the phase diagram. Collectively, these studies reveal a common observation: under constant pressure or external fields, the free energy of a system does not necessarily vary monotonically with temperature, but may reverse multiple times due to changes in vibrational entropy, magnetic entropy, volume effects, or molecular rearrangements, thereby giving rise to re-entrant phase transitions.

Against this backdrop, we performed high-precision quasiharmonic free-energy calculations for high-pressure phases of $Ta_2O_5$. The results show that at $P$ = 2 GPa (Fig. 7), the entropy and free energy difference $G_B(T)$-$G_\gamma(T)$ exhibits a pronounced non-monotonic temperature dependence (see Figs. 7(c), 7(d)). At around 60 K, the free energies of the two phases intersect for the first time, making the B phase thermodynamically more stable. As the temperature increases further, competition involving vibrational entropy again alters the free-energy ordering, leading to a second crossing near 208 K, where the γ phase regains stability—constituting a typical re-entrant sequence.

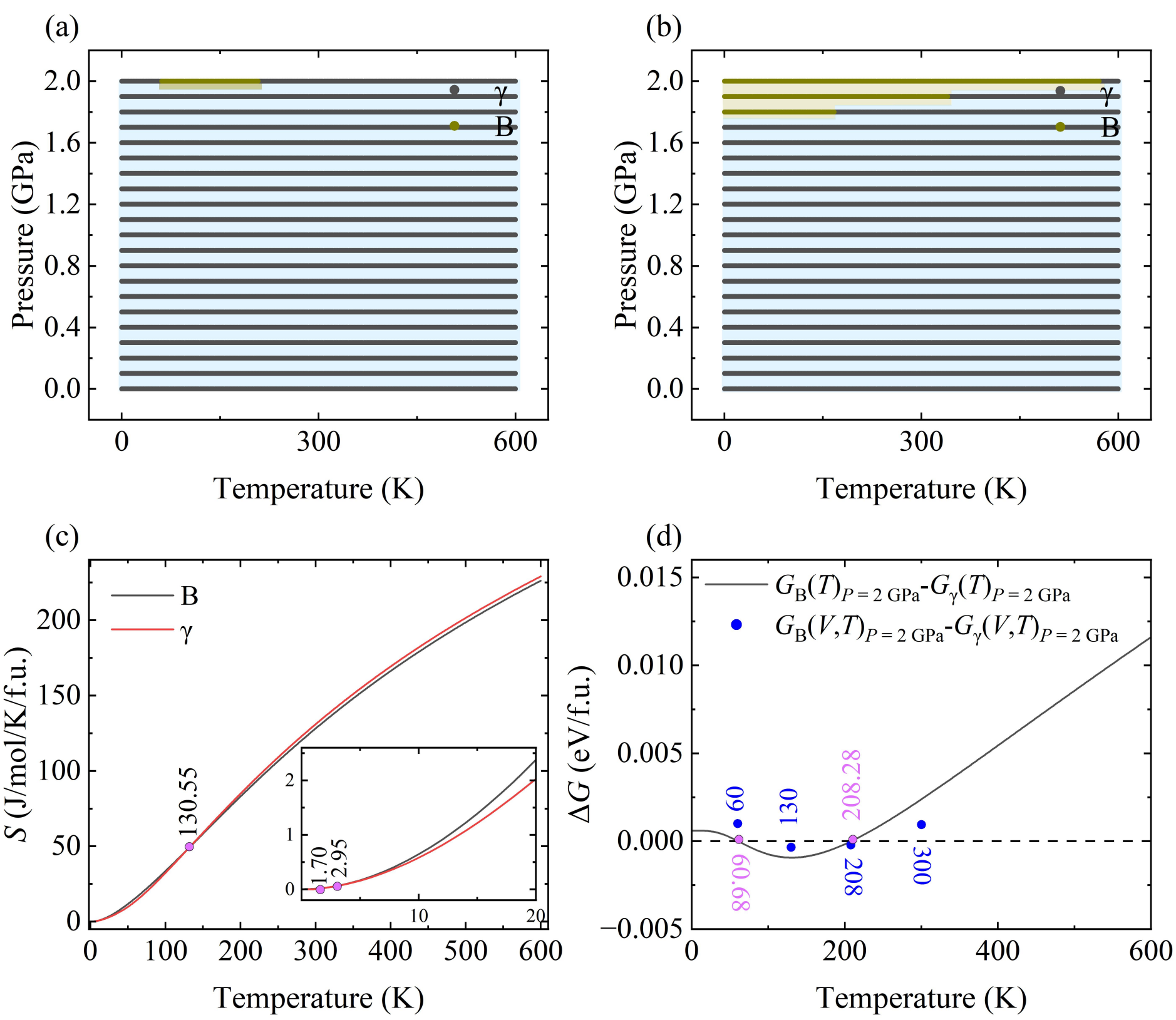

**FIG. 7.** The phase diagram of γ and B-$Ta_2O_5$, calculated at a pressure step of 0.1 GPa and temperature step of 1 K, for the case with (panel a) and without (panel b) phonon energies. Temperature dependence of entropy (panel c) and free energy difference (panel d) of γ and B-$Ta_2O_5$. The scattered data points displayed in panel (d) correspond to $\Delta G(T)$ calculated within QHA, which includes thermal expansion effects.

Within the quasi-harmonic approximation (QHA) [82-84], the effects of volume variation due to temperature can be evaluated. The structure at $P \sim 2$ GPa ($T \sim 0$ K) is taken as the reference configuration. The enthalpy is calculated at slightly expanded and contracted unit cell volumes, thereby establishing the enthalpy–volume relation. Subsequently, DFPT calculations are carried out to obtain the vibrational free energy at each volume *V*, allowing the construction of Gibbs free energy as a function of *P* and *T*. The equilibrium volume at a given *P* and *T* is determined by minimizing the free energy with respect to volume: $G(P,T) = \min_{V} \left[U(V) + F_{ph}(V,T) + PV\right]$, where the terms $U$ and $F_{ph}$ have the same meaning as above. The temperature-volume (*T*–*V*) relationship is then established with the effects of thermal expansion incorporated.

Using the *T*–*V* data, calculations are performed at thermally expanded volumes corresponding to *T* = 60, 130, 208, and 300 K to obtain the Gibbs free energies of γ and B-$Ta_2O_5$ at *P* = 2 GPa,.from which the difference ($\Delta G$) is obtained. The results shown in Fig. 7(d) indicate that subtle variations in volume can lead to noticeable corrections and slightly shift the crossing points of $\Delta G$ between the two phases. Nevertheless, the overall trend of re-entrant phase transition remains unchanged.

To further examine the finite-temperature stability of γ and B-$Ta_2O_5$, AIMD simulations were performed, which intrinsically include the anharmonic effects of atomic vibrations. As shown in Fig. 8, the enthalpies (*H*) of both γ and B-$Ta_2O_5$ at *T* = 10 K, 60 K, 130 K, and 300 K exhibit regular (nearly equal-magnitude) fluctuations around an average value (e.g., *T* = 60 K, $\overline{H}$ = −67.1246 eV/f.u. and −67.1236 eV/f.u., for the γ and B phase, respectively) without noticeable drift or divergence, indicating that their dynamical stability under thermal perturbations.

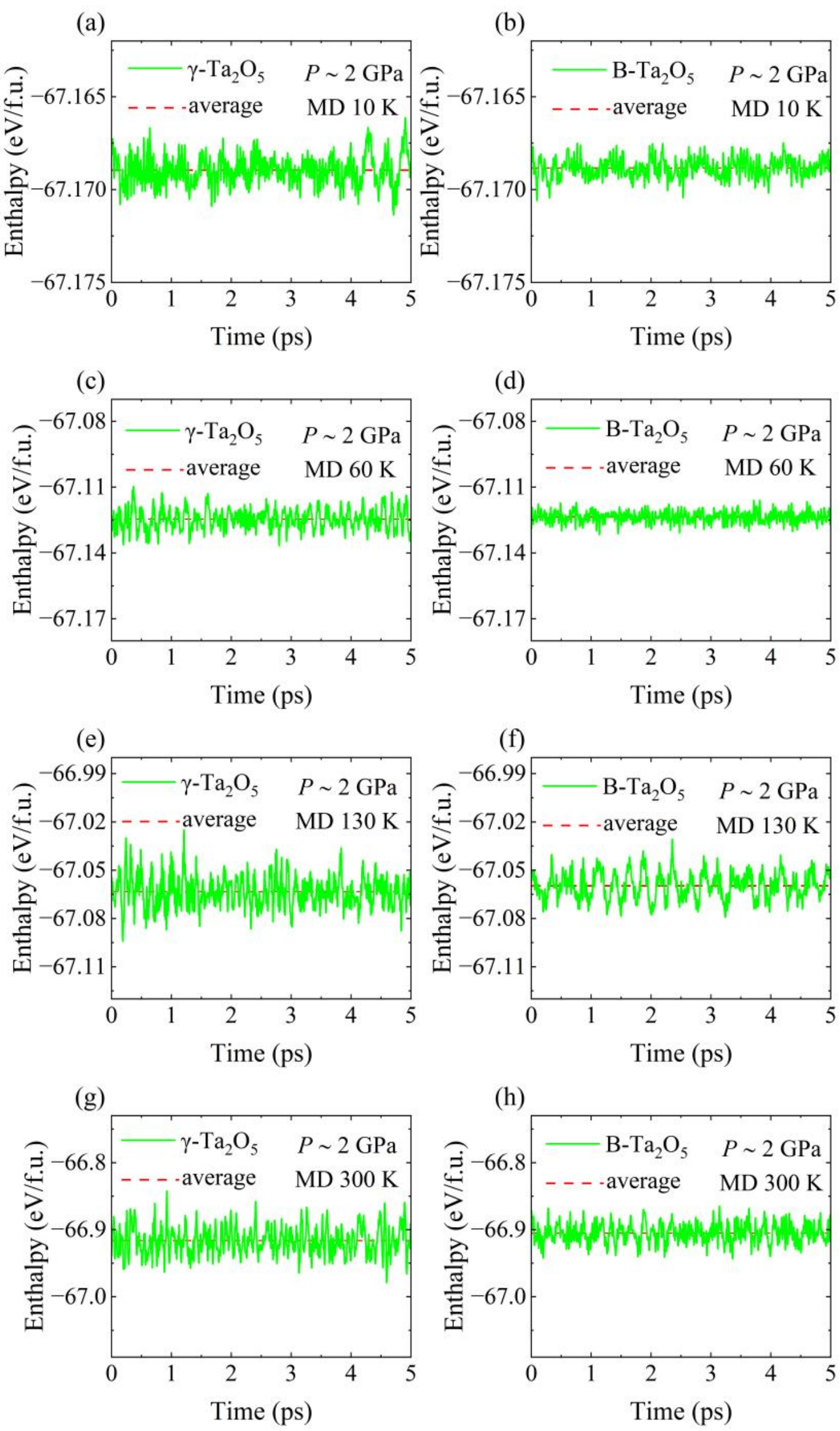

**FIG. 8**. Time evolution of the enthalpies of γ and B-$Ta_2O_5$ by AIMD, for $T$ = 10 K (a, b), 60 K (c, d), 130 K (e, f), and 300 K (g, h),at $P$ ~ 2 GPa.

As shown in Fig. 9(a), at $P$ ~ 2 GPa, the vibrational entropy difference between B and γ phases, $\Delta S(T) = S_B - S_\gamma$, exhibits a pronounced non-monotonic temperature dependence. In the low-temperature regime (0–130 K), $\Delta S > 0$, indicating that the B phase has a higher vibrational entropy. With increasing temperature, $\Delta S$ gradually decreases and approaches zero at $T \approx 130$ K, marking a critical point where the entropy difference changes sign. Using the time-averaged enthalpy ($\overline{H}$) obtained by AIMD simulations, the time-averaged Gibbs free energy difference can be deduced as $\Delta\overline{G}(T) = \Delta\overline{H} - T\Delta S$. The AIMD results indicate that B phase has a slightly higher enthalpy than the γ phase ($\Delta\overline{H} > 0$). As shown in Fig. 9(b), at low temperatures, the negative entropy contribution compensates for the enthalpy difference, making $\Delta\overline{G}$ close to zero or slightly negative (e.g., $T$ = 60 K, $\Delta\overline{G} \sim -0.33$ meV/f.u.). However, above 130 K where $\Delta S < 0$, the term $-T\Delta S$ together with the enthalpy contribution leads to an increase in $\Delta\overline{G}$. For instance, at $T$ = 300 K, $\Delta\overline{G} \approx 0.02$ eV/f.u., indicating that the γ phase is thermodynamically more favored. The above results further demonstrate the existence of possible re-entrant phase transition between the B and γ phase of $Ta_2O_5$ from a thermodynamic perspective.

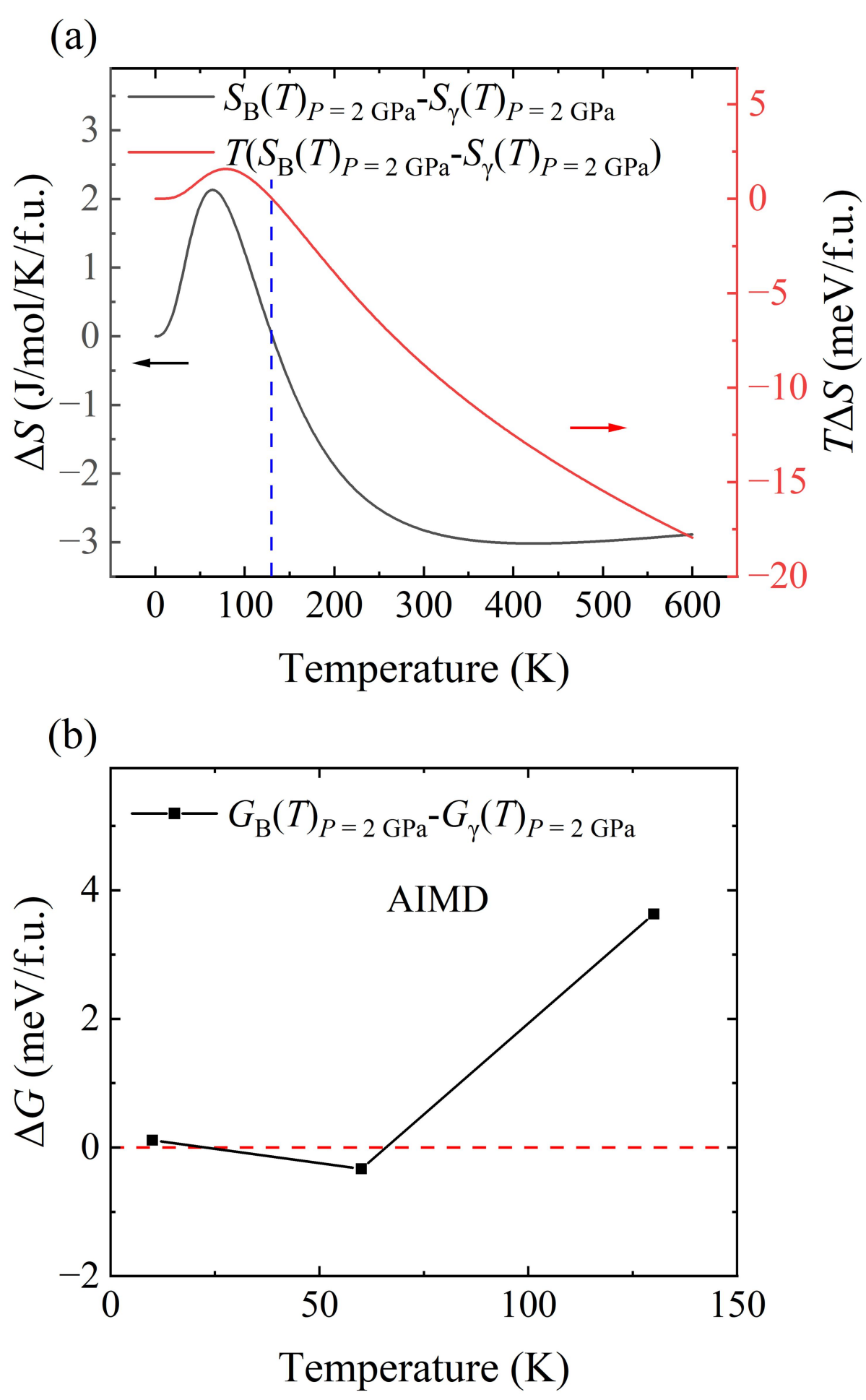

**FIG. 9**. Temperature-dependent thermodynamic differences between B and γ phases at $P \sim 2$ GPa. (a) Vibrational entropy difference $\Delta S$ and its contribution $T\Delta S$. The blue dash line indicates the temperature at which $\Delta S = 0$. (b) Time-averaged free energy difference $\Delta\overline{G}(T)$ from AIMD simulations.

### C. The Zero-Temperature Phase Transition Pressure

At absolute zero ($T$ = 0 K), the enthalpy of each phase is calculated as $H = E + PV$. A pressure-induced structural phase transition occurs when the enthalpies of two phases equal, with the corresponding pressure defined as the critical pressure ($P_c$).

$$P_c = (E_{tot2} - E_{tot1})/(V_1 - V_2) = -\Delta E/\Delta V, \quad (3)$$

where $E_{tot1}$ and $E_{tot2}$ represent the internal energies of two crystal structures, while $V_1$ and $V_2$ correspond to the volumes of these respective structures. Two situations were considered: (i) without zero-point energy (ZPE) correction, where $E_{tot} = E_0$, where $E_0$ is the ground-state energy; (ii) with ZPE correction, where $E_{tot} = E_0 +$ ZPE. The resulting critical pressures ($P_{c1}$ (GPa) and $P_{c2}$ (GPa, respectively) are summarized in Table **III**. The discrepancies between $P_{c1}$ and $P_{c2}$ demonstrate that ZPE exerts a non-negligible effect on phase transition pressures near criticality.

For example, the γ→B transition pressure increases from 1.78 GPa (without ZPE) to 2.56 GPa (with ZPE), while the $\gamma_1$→B transition shifts from 0.67 GPa to 1.48 GPa. These results demonstrate that incorporating the zero-point energy correction leads to nontrivial changes in the transition pressures, underscoring the importance of NQEs in accurately describing $Ta_2O_5$ phase stability.

Understanding the critical pressures at 0 K is essential for revealing how materials behave under extreme conditions. For $Ta_2O_5$, the γ phases are more favorable at low pressures, while the B phase becomes increasingly stable at high pressures. The γ→B and $\gamma_1$→B transition pressures further indicate that γ phases are more suitable for applications requiring stability at low pressures, whereas the B phase exhibits enhanced stability at high pressures. Furthermore, it can be seen from Table V that the disparity between $P_{c1}$ and $P_{c2}$, $\Delta P$ (= $P_{c2}$ - $P_{c1}$), for the transition from γ and $\gamma_1$ to the other polymorphs (e.g., B, λ, $L_{SR}$, ...) is very similar. This is due to the fact that the ground state energy and crystal volume per formula unit (mass density) of γ and $\gamma_1$-$Ta_2O_5$ is very similar [43].

**Table III.** Summary of critical pressures for various $Ta_2O_5$ phases, with $\Delta P = P_{c2} - P_{c1}$.

| $Ta_2O_5$ | γ, B | γ, λ | γ, $L_{SR}$ | γ, $β_R$ | γ, Z | $β_R$, Z | λ, Z | $L_{SR}$, Z |
|---|---|---|---|---|---|---|---|---|
| $P_{c1}$ (GPa) | 1.78 | 4.09 | 6.25 | 6.68 | 19.84 | 28.61 | 31.37 | 27.61 |
| $P_{c2}$ (GPa) | 2.56 | 4.76 | 6.20 | 6.59 | 20.43 | 29.06 | 31.72 | 28.08 |
| Δ$P$ (GPa) | 0.78 | 0.67 | -0.05 | -0.09 | 0.59 | 0.45 | 0.35 | 0.47 |
| $Ta_2O_5$ | $γ_1$, B | $γ_1$, λ | $γ_1$, $L_{SR}$ | $γ_1$, $β_R$ | $γ_1$, Z | $β_{AL}$, Z | δ, Z | $γ_1$, δ |
| $P_{c1}$ (GPa) | 0.67 | 2.74 | 4.61 | 5.21 | 19.34 | 6.09 | 11.73 | 39.16 |
| $P_{c2}$ (GPa) | 1.48 | 3.37 | 4.47 | 5.15 | 19.94 | 7.42 | 12.53 | 38.99 |
| Δ$P$ (GPa) | 0.81 | 0.63 | -0.14 | -0.06 | 0.60 | 1.33 | 0.80 | -0.17 |

### D. Role of Zero-Point Energy (ZPE) at Finite Temperatures

The analysis above comparing systems with and without phonon energy contributions quantitatively demonstrated that nuclear quantum effects (NQEs) significantly modify phase boundaries and stability sequences. We now systematically examine how zero-point energy (ZPE) contributes to crystal phonon energetics and its critical role in lattice dynamics and thermodynamic stability.

The phonon energy at a finite temperature $T$ reads:

$$E_{ph} = Re[\sum_j \left(\frac{1}{2}\hbar\omega_j + \frac{\hbar\omega_j}{e^{\hbar\omega_j/(k_B T)} - 1}\right)] = Re[\sum_j \left(\frac{1}{2}\hbar\omega_j\right)] + Re[\sum_j \left(\frac{\hbar\omega_j}{e^{\hbar\omega_j/(k_B T)} - 1}\right)]$$
$$= ZPE + E_{ph,T>0} \quad (4)$$

Only the modes with real vibrational frequencies are considered in the summation. In addition to the zero-point energy (ZPE), the phonon energy at non-zero temperature ($E_{ph,T>0}$) may be expressed as follows:

$$E_{ph,T>0} = Re[\sum_j \left(\frac{\hbar\omega_j}{e^{\hbar\omega_j/(k_B T)} - 1}\right)] = \int_0^{\omega_m} \left[\frac{\hbar\omega}{e^{\hbar\omega/(k_B T)} - 1}\right] g(\omega) d\omega, \quad (5)$$

where $\omega_m$ denotes the maximum frequency and $g(\omega)$ is the vibrational density of states. Using Eq. (5), we can calculate the energy at non-zero temperature $E_{ph,T>0}$, then derive the total phonon energy $E_{ph}$ using Eq. (4). The ZPE term and $g(\omega)$ were readily obtained using DFPT calculations. We further determined its relative contribution to the total phonon energy (Fig. 10). The results show that ZPE constitutes the major component at cryogenic temperatures and its proportion progressively drops with increasing $T$, consistent with the growing dominance of thermally activated lattice vibrations. Moreover, for a given temperature, the ZPE fraction increases systematically with pressure, indicating that compression enhances NQEs.

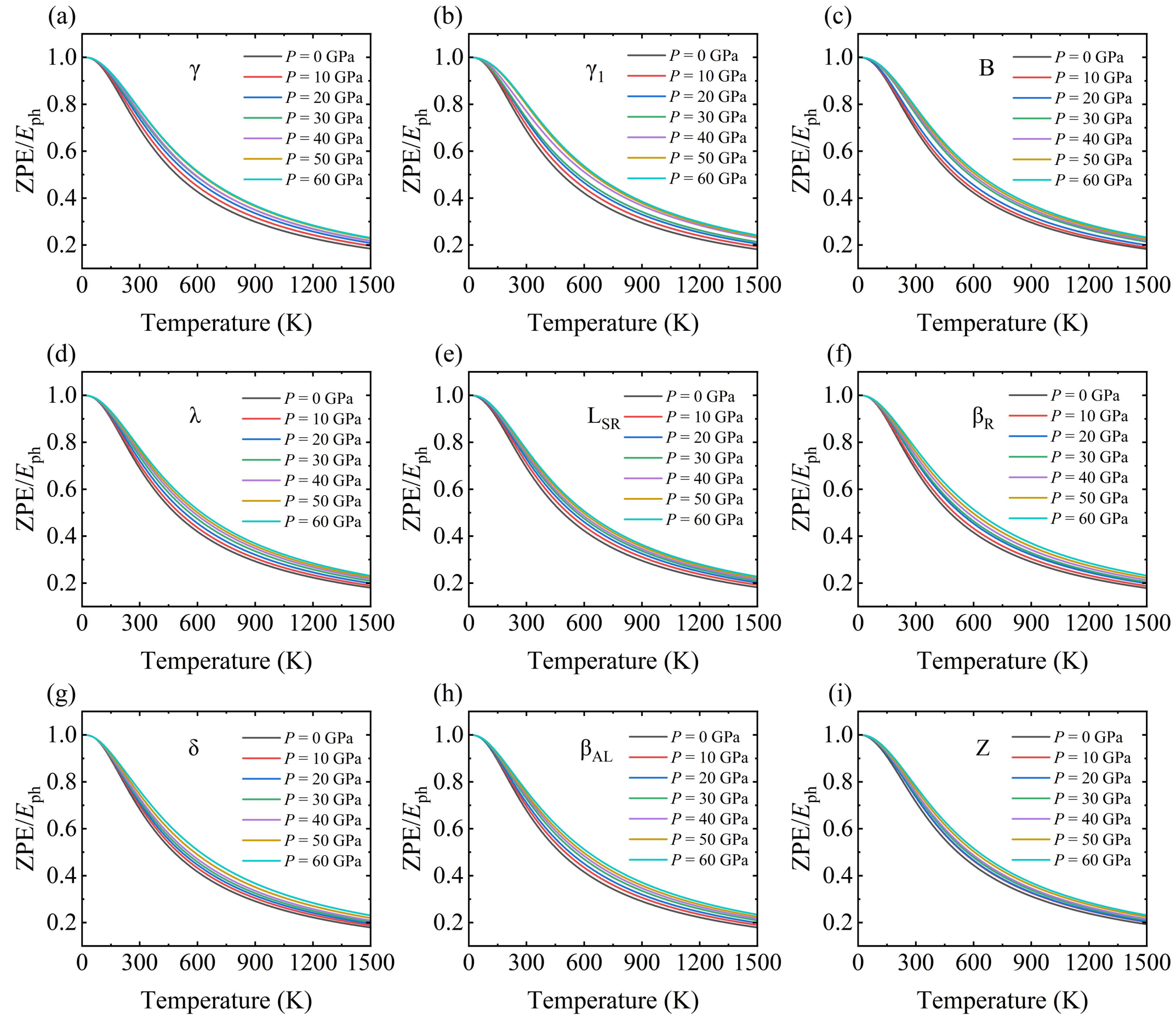

**FIG. 10.** Temperature dependence of the zero-point energy (ZPE) contribution to the total phonon energy $E_{ph}$ of nine $Ta_2O_5$ polymorphs at different pressures.

We further investigated the role of ZPE in lattice dynamics by analyzing the ratio between zero-point energy (ZPE) and nonzero-temperature phonon energy $E_{ph,T>0}$, identifying a characteristic temperature $T_0$ where their contributions become equal (i.e., ZPE/$E_{ph,T>0}$ = 1). As shown in Fig. 11, ZPE dominates the vibrational spectrum at cryogenic temperatures. Its role gradually decreases with increasing temperature and reaches the critical point $T_0$, at which the zero-point and thermal motions of atoms contribute equally to the total phonon energies. The precisely determined $T_0$ serves as a fundamental parameter that quantifies the temperature range where quantum effects govern lattice behavior. These results establish $T_0$ as both a crucial metric for evaluating computational methods and a practical design parameter for quantum material applications. Concurrently, it can be seen that the contribution of ZPE increases with pressure at a given temperature

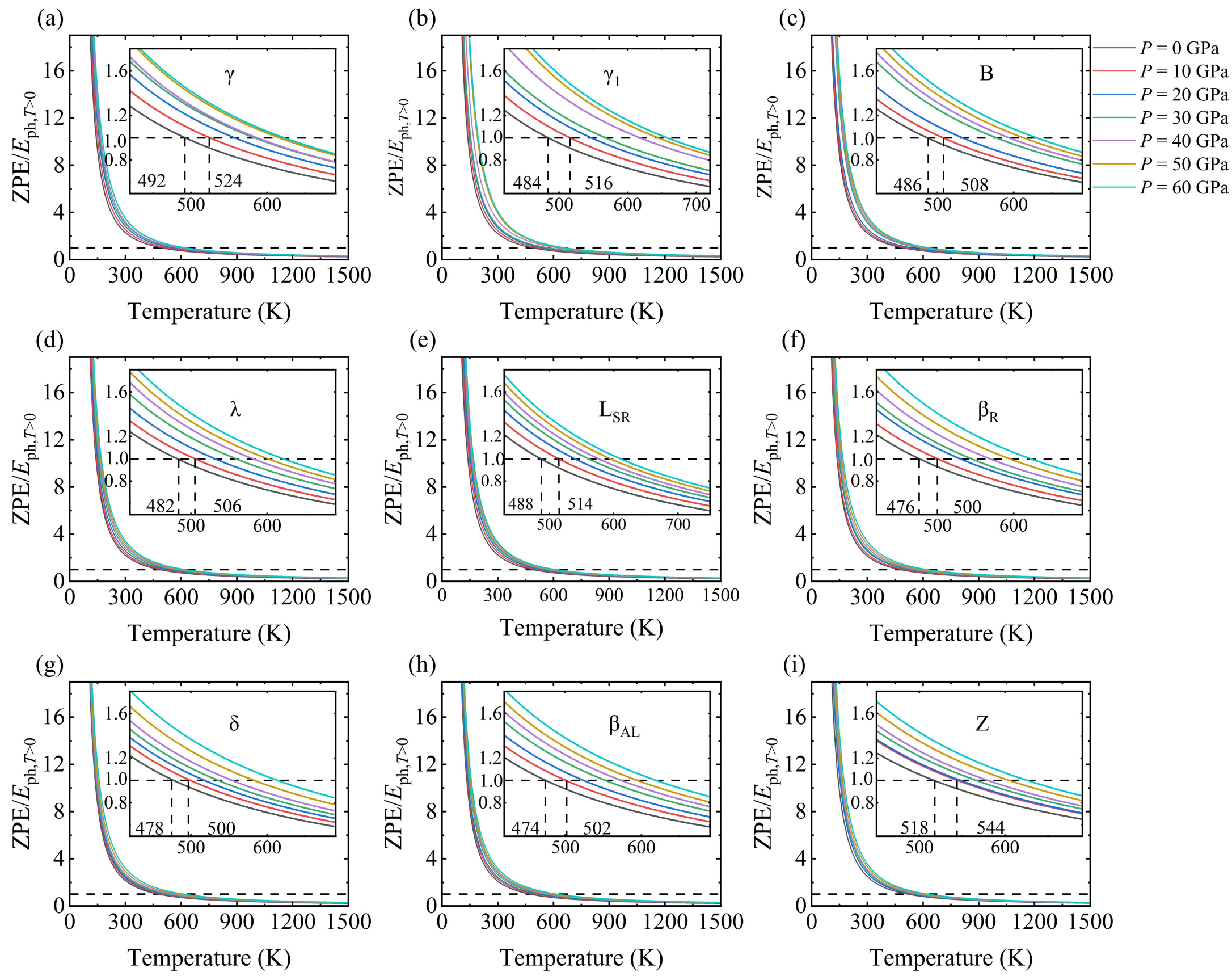

**FIG. 11.** Temperature dependence of the ratio between zero-point energy (ZPE) and nonzero-temperature phonon energy ($E_{\mathrm{ph,T>0}}$) of nine $Ta_2O_5$ polymorphs at different pressures. The characteristic temperatures $T_0$ (where ZPE/$E_{\mathrm{ph,T>0}}$ = 1) are explicitly indicated (horizontal dashed lines).

Figure 12(a) presents the pressure dependence of the characteristic temperatures $T_0$ and their average of the nine $Ta_2O_5$ phases. Despite some fluctuations, the overall trend that pressure enhances $T_0$ and the role of ZPE in a given system is found. As shown by two of the authors [85], there exists a characteristic temperature (~ 1/3 of the Debye temperature $T_D$) at which the atomic mean square displacement (MSD) due to zero-point and thermal motions are equal. To correlate these observations, we have further derived the maximum phonon frequency ($\nu_{max}$) for each polymorph from DFPT calculations and estimated the corresponding Debye temperature ($T_D \sim h\nu_{max}/k_B$), where h is the Planck constant and $k_B$ the Boltzmann constant. Shown in Fig. 12(b), are the calculated $T_D/T_0$ values for the nine $Ta_2O_5$ polymorphs with varying pressures. Within a broad range of pressure, the number $T_D/T_0$ is nearly constant for each polymorph, which is roughly described by $T_D/T_0 = \mu \sim (3 \pm 0.5)$. In particular, four phases ($\gamma$, $\gamma_1$, $\delta$, $\beta_R$) exhibit excellent agreement with the relation of $T_D/T_0 = 3$, with discrepancies smaller than 3%. This coincidence implies that the Debye model accounts for the phonon properties of these $Ta_2O_5$ phases ($\gamma$, $\gamma_1$, $\delta$, $\beta_R$). Considering the fact that the vibrational frequency is proportinal to the inverse of MSD [85] (equivalently, the phonon energy), the characteristic temperature which yields equal MSDs would give approximately the same energies for both zero-point and thermal phonons. These results provide preliminary support for extension of the earlier theoretical prediction regarding the relationship between the Debye temperature and the quantum-to-thermal crossover point [85], from elemental crystal systems to much more complex transition metal oxides.

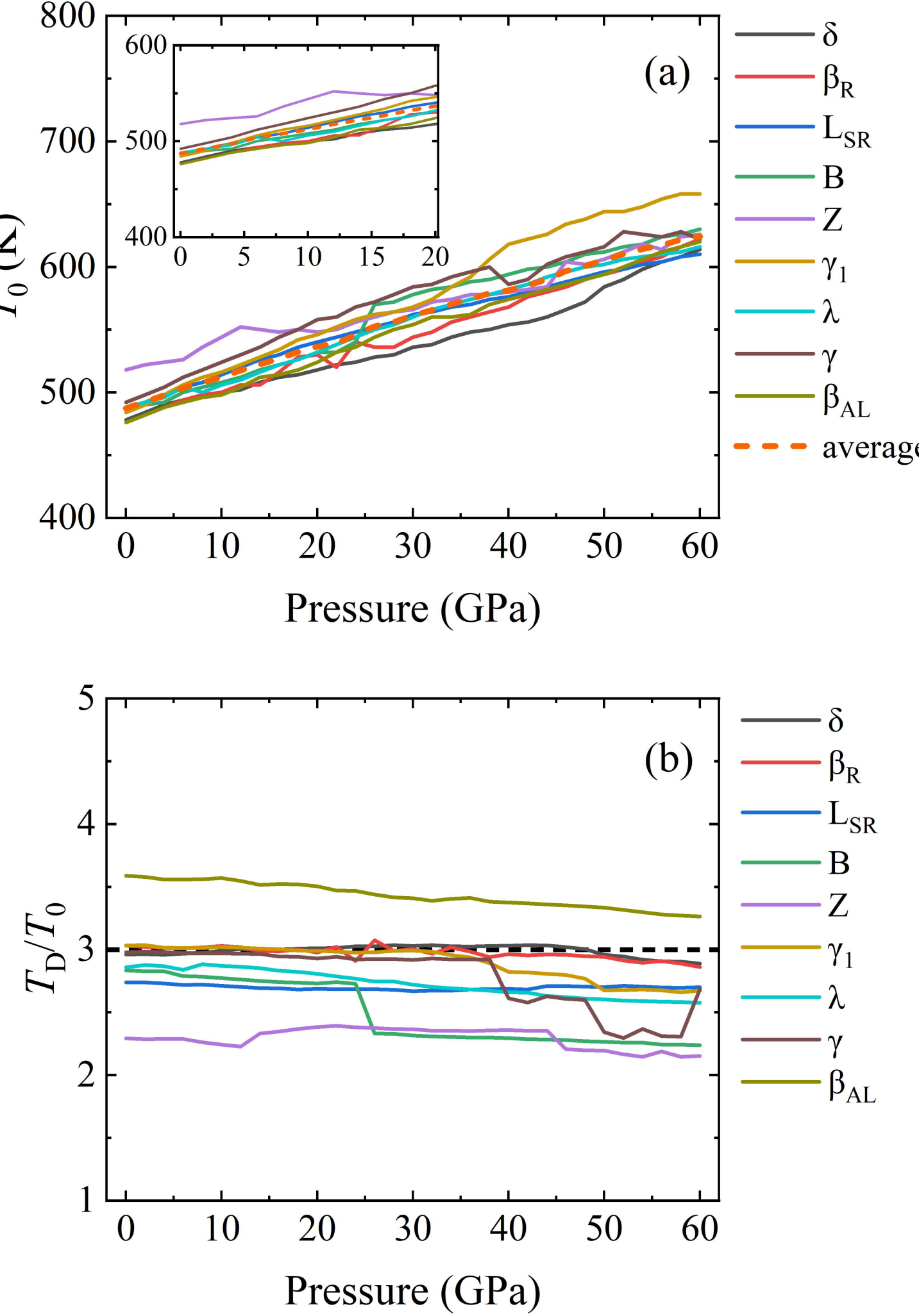

**FIG. 12.** Pressure dependence of the characteristic temperature $T_0$ (panel a) and its ratio to the Debye temperature ($T_D/T_0$, panel b), for the $Ta_2O_5$ polymorphs studied above.

## IV. Conclusions

In summary, this study establishes a comprehensive phase diagram of $Ta_2O_5$ across a broad temperature and pressure range using first-principles calculations. The γ-$Ta_2O_5$ and B-$Ta_2O_5$ are identified as two primary stable phases, with γ-$Ta_2O_5$ dominating at low temperature and pressures, and B-$Ta_2O_5$ stabilized under high-temperature and high-pressure conditions. The precisely clarified stability ranges of γ-$Ta_2O_5$ and B-$Ta_2O_5$, including their coexistence boundaries and metastable regimes, provide crucial theoretical guidance for targeted synthesis strategies, particularly for technological applications requiring structural robustness under ambient conditions versus extreme environments. Importantly, nuclear quantum effects (NQEs), manifestations of the quantum motions of atomic nuclei, are shown to significantly influence the phase stability sequence, highlighting their importance even in heavy-element systems such as transition metal oxides. The re-entrant phase transition between γ and B-$Ta_2O_5$ at $P \sim 2$ GPa is predicted, a demonstration of the nontrivial role of NQEs in the thermodynamics of this system. In addition, it is shown that subtle volume variations arising from thermal expansion can result in a slight shift in the crossing point of $\Delta G$ between the two phases. Furthermore, *ab initio* molecular dynamics (AIMD) simulations confirm the dynamical stability of both the γ and B phases near the transition region, and validate the re-entrant phase transition between the γ and B phases driven by enthalpy–entropy competition. Another finding is the determination of characteristic temperature $T_0$ (approximately one-third of the Debye temperature), which physically corresponds to the crossover point where zero-point vibrations balance thermal phonon contributions to the free energy. Looking ahead, more advanced simulations that incorporate larger supercells and anharmonic effects, together with experimental validation, will be essential to further refine the stability map and assess the potential of wider range of applications.

**Acknowledgements**

This work is financially supported by the National Natural Science Foundation of China (No. 12074382, 11474285, 12464012). We are grateful to the staff of the Hefei Branch of Supercomputing Center of Chinese Academy of Sciences, and the Hefei Advanced Computing Center for support of supercomputing facilities. We would like to thank the crew of the Center for Computational Materials Science, Institute for Materials Research of Tohoku University, and the supercomputer resources through the HPCI System Research Project (hp200246). We also thank Yangwu Tong for helpful discussions.

DATA AVAILABILITY

The data that support the findings of this article are available within the article and the Science Data Bank [86].